%% file: main.tex
\documentclass[11pt]{article}

\usepackage[final]{acl}

\usepackage{times}
\usepackage{latexsym}
\usepackage{booktabs}
\usepackage{amsmath}

\usepackage[T1]{fontenc}

\usepackage[utf8]{inputenc}

\usepackage{microtype}
\usepackage{marvosym}

\usepackage{inconsolata}

\usepackage{graphicx}
\usepackage{multirow}
\usepackage[table,xcdraw]{xcolor}
\usepackage{longtable}
\usepackage{textcomp}   

\title{Unlocking the Edge deployment and ondevice acceleration of multi-LoRA enabled one-for-all foundational LLM}

\author{
\textbf{Sravanth Kodavanti}\textsuperscript{$\dagger,*$, \Letter},\;
\textbf{Sowmya Vajrala}\textsuperscript{$\dagger,*$},\;
\textbf{Srinivas Miriyala}\textsuperscript{$\dagger,*$} \\[2pt]
\textbf{Utsav Tiwari}\textsuperscript{$\dagger$},\;
\textbf{Uttam Kumar}\textsuperscript{$\dagger$},\;
\textbf{Utkarsh Kumar Mahawar}\textsuperscript{$\dagger$},\;
\textbf{Achal Pratap Singh}\textsuperscript{$\dagger$} \\[2pt]
\textbf{Arya D}\textsuperscript{$\dagger$},\;
\textbf{Narendra Mutyala}\textsuperscript{$\dagger$},\;
\textbf{Vikram Nelvoy Rajendiran}\textsuperscript{$\dagger$},\;
\textbf{Sharan Kumar Allur}\textsuperscript{$\dagger$} \\[2pt]
\textbf{Euntaik Lee}\textsuperscript{$\ddagger$},\;
\textbf{Dohyoung Kim}\textsuperscript{$\ddagger$},\;
\textbf{HyeonSu Lee}\textsuperscript{$\ddagger$},\;
\textbf{Gyusung Cho}\textsuperscript{$\ddagger$},\;
\textbf{JungBae Kim}\textsuperscript{$\ddagger$} \\[2pt]
\textsuperscript{$\dagger$} Samsung Research Institute Bangalore, India, 
\textsuperscript{$\ddagger$} Samsung Electronics, Suwon, South Korea \\[2pt]
\textsuperscript{*} Equal Contribution \\
\textsuperscript{\Letter} Corresponding author: k.sravanth@samsung.com
}

\begin{document}
\maketitle
\input{section/0_abstract}
\input{section/1_introduction}
\input{section/2_reported_works}
\input{section/3_proposed_method}
\input{section/4_experiments_methods}
\input{section/5_conclusion}
\input{section/6_limitations}
\bibliography{custom}
\input{section/7_appendix}

\end{document}

%% file: section/0_abstract.tex
\begin{abstract}
Deploying large language models (LLMs) on smartphones poses significant engineering challenges due to stringent constraints on memory, latency, and runtime flexibility. In this work, we present a hardware-aware framework for efficient on-device inference of a LLaMA-based multilingual foundation model supporting multiple use cases on Samsung Galaxy S24 and S25 devices with SM8650 and SM8750 Qualcomm chipsets respectively. Our approach integrates application-specific LoRAs as runtime inputs to a single frozen inference graph, enabling dynamic task switching without recompilation or memory overhead. We further introduce a multi-stream decoding mechanism that concurrently generates stylistic variations - such as formal, polite, or jovial responses - within a single forward pass, reducing latency by up to 6×. To accelerate token generation, we apply Dynamic Self-Speculative Decoding (DS2D), a tree-based strategy that predicts future tokens without requiring a draft model, yielding up to 2.3× speedup in decode time. Combined with quantization to INT4 and architecture-level optimizations, our system achieves 4-6× overall improvements in memory and latency while maintaining accuracy across 9 languages and 8 tasks. These results demonstrate practical feasibility of deploying multi-use-case LLMs on edge devices, advancing the commercial viability of Generative AI in mobile platforms.
\end{abstract}

%% file: section/1_introduction.tex
\section{Introduction}
Deploying Large Language Models (LLMs) on mobile devices promises significant benefits in privacy, latency, and offline accessibility. However, bringing the flexibility of server-scale LLM adaptation-especially via Parameter-Efficient Fine-Tuning (PEFT) methods like LoRA-into the constrained environment of on-device inference presents several non-trivial challenges. 

Unlike server-side systems, where models can be fine-tuned, recompiled, and dynamically loaded, on-device deployment must operate with frozen inference graphs, minimal memory, and strict runtime performance guarantees. Bridging the disconnect between flexible model development and the immutability of on-device inference demands a fundamental shift in how adaptable LLMs are engineered for real-world deployment.

In this work, we present a practical framework for real-time, multilingual, multi-use case LLM inference on Qualcomm's Neural Processing Unit (NPU) on SM8650 and SM8750 chipsets deployed on commercial smartphones, specifically the Samsung Galaxy S24 and S25. Our solution builds on a quantized LLaMA-based model integrated with eight application-specific LoRA modules across nine languages. Unlike conventional LoRA usage, where models are statically merged during training, we treat LoRAs as runtime inputs to the frozen inference graph, enabling dynamic task switching without recompilation or re-quantization. This design significantly reduces storage requirements and supports plug-and-play use cases on embedded device.

To meet hardware efficiency constraints, we perform targeted architectural transformations, including converting multi-head attention into parallel single-head paths, reparameterizing linear layers into convolutions, and quantizing both activations and weights to 4-bit precision (INT4) while utilizing mixed precision strategies to balance accuracy. These optimizations achieve substantial improvements in memory and latency while maintaining accuracy within acceptable limits.

Beyond architectural and quantization enhancements, we introduce a multi-stream token generation strategy that enables simultaneous inference across stylistic variants within a single decoding pass. For use cases such as tone transfer or stylistic rewriting (e.g., converting input into polite, formal, or jovial responses), traditional inference would require eight separate decoding iterations-one per style. By exploiting the fact that all use cases share the same frozen graph and memory layout, and that stylistic variants often are driven by the first token, we design a masked decoding scheme that modifies the first-token sampling and allows concurrent generation of eight distinct outputs over a shared KV-cache and tensor layout. This yields up to 6× latency and memory reduction for stylistic generation without any change to the model binary or graph.

Finally, to address decode-time latency bottlenecks, we propose a speculative sampling strategy based on tree-based branching with prefix tuning. Our method operates without needing a separate draft model or retraining, making it fully compatible with frozen, single-graph inference pipelines. This enables semi-autoregressive decoding with up to 2× improvement in tokens-per-second throughput, further pushing the boundary of what’s achievable in mobile LLM deployment without reliance on cloud infrastructure.

%% file: section/2_reported_works.tex
\section{Reported Works}
The Transformer architecture has played a pivotal role in the advancement of generative AI, enabling large language models (LLMs) like LLAMA \cite{touvron2023llama}, GPT \cite{achiam2023gpt} to produce high-quality, human-like text and opening up new possibilities for applications such as language translation, text summarization, and content generation. However, the increasing size and complexity of these models have also raised concerns about their computational efficiency and environmental sustainability. To address these challenges, researchers have proposed various techniques to optimize LLMs for deployment on resource-constrained devices.

Recent works have focused on developing efficient LLMs that can be deployed on mobile and edge devices.  MobiLlama \cite{thawakar2024mobillama} is one such model that has gained significant attention due to its compact size and competitive performance. 


Quantization, on the other hand, is a widely used technique to reduce the precision of model weights and activations, thereby reducing memory footprint and computational requirements. QLoRA \cite{dettmers2023qlora} and QaLoRA \cite{xu2023qa} quantization-based low-rank adaptation  techniques that have shown promising results in reducing the precision of LLMs while maintaining their performance. 

Low-Rank Adaptation (LoRA) \citep{hu2022lora} is a technique used to adapt large language models (LLMs) to specific tasks or domains by applying low-rank matrices to the model weights. Quantization, on the other hand, is a widely used technique to reduce the precision of model weights and activations, thereby reducing memory footprint and computational requirements. QLoRA \cite{dettmers2023qlora} and QaLoRA \cite{xu2023qa} are two recent quantization-based low-rank adaptation techniques that have shown promising results in reducing the precision of LLMs while maintaining their performance.

These techniques have been applied to various NLP tasks, including language modeling and text classification.

In addition to quantization, model optimizations like Flash Attention \cite{dao2205fast} have been proposed to reduce the computational overhead of self-attention mechanisms in transformers. Flash Attention achieves this by using a novel attention mechanism that reduces the number of computations required.


Apart from efficient attention, Speculative decoding is another prominent technique used to accelerate the inference speed of LLMs by predicting the output tokens in parallel. Eagle \cite{li2024eagle}, Medusa \cite{cai2024medusa}, and BiTA \cite{lin2025bita} are recent works that have proposed novel speculative decoding methods to improve the inference speed of LLMs.

Despite the advancements in efficient large language models (LLMs), quantization techniques, and speculative decoding methods, deploying a single LLM for multiple tasks on resource-constrained devices remains a significant challenge. Even efficient models like  MobiLlama are not sufficient to handle multiple tasks across different languages. Furthermore, optimizations like Flash Attention are not compatible with Neural Processing Units (NPUs), and draft-based speculative decoding methods require additional memory, which is a scarce resource on constrained devices. It is essential to develop a one-for-all foundation model along with effective optimization methods to facilitate its deployment on resource-constrained devices.

%% file: section/3_proposed_method.tex
\begin{figure*}[t]
\centering
\includegraphics[width=1\textwidth, height=7.5cm]{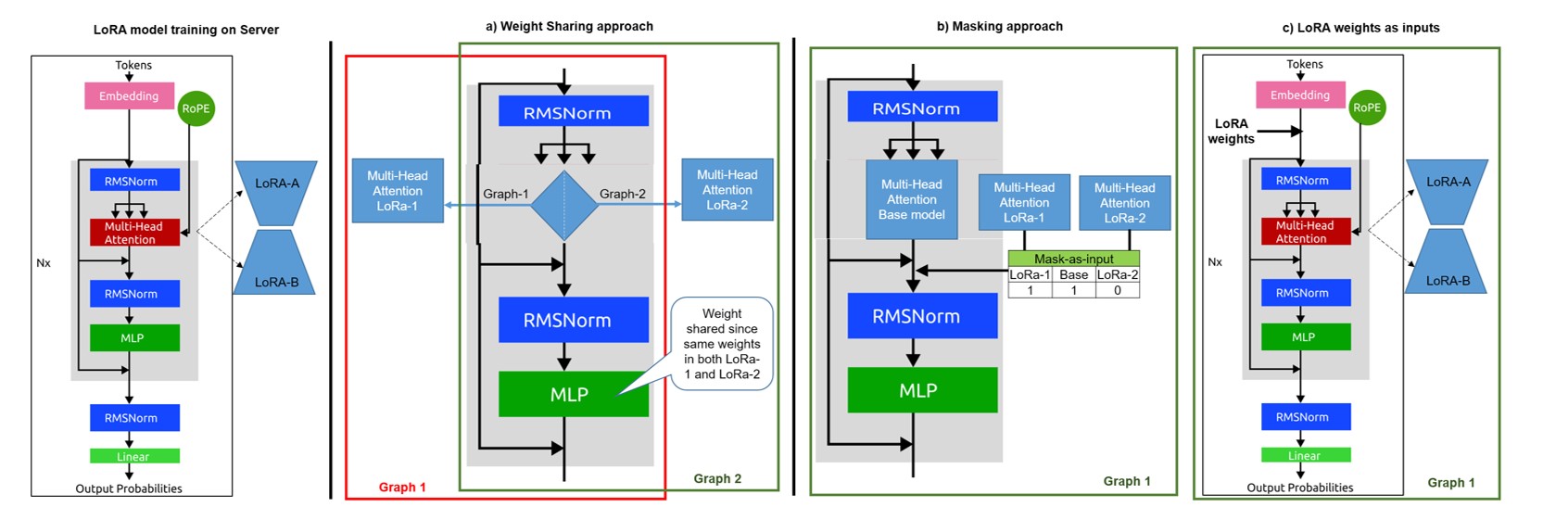}  
\captionsetup{justification=centering}
\caption{Proposed method for enabling multiple LoRA with a single LLM on embedded device.}
\label{fig_lora}
\end{figure*}

\section{Proposed Method}
This work outlines the process starting from a trained foundation model with LoRAs, optimizing and deploying it on smartphones for no-cloud, fully on-device execution. It focuses on multi-LoRA enablement, inference optimizations, a novel lossless strategy for concurrent token generation, and speculative decoding for decode time acceleration. The authors refrain from describing the details of training the proprietary foundation model and LoRAs and solely focus on the innovations curated to ensure the on-device deployment. The readers are encouraged to refer to the seminal works on LLAMA \citep{touvron2023llama} and LoRA \citep{hu2022lora}, which inspired the authors to build their foundation models.  

\subsection{LLM architecture and LoRA updates}

The authors experiment with 1-billion and 3-billion parameter LLAMA-based LLM models, employing a decoder-only architecture with D sequential decoders. Tokenization generates embeddings of size E, which are transformed into Query (Q), Key (K), and Value (V) through linear layers, RMS normalization, and Rotary Position Embedding (RoPE) applied to Q and K. Each decoder block includes a multi-head attention (MHA) sub-block with H heads, followed by RMS normalization and an MLP block. The MHA output map (O) is processed by a learnable linear layer, producing Q, K, V, and O projection matrices with dimensions E x L and L x E, respectively, where L is the latent dimension.

In this work, LoRA weight updates are applied to these projection matrices (Eqs. 1–4), targeting the attention block. For each LoRA-supported layer, two parallel layers—LoRA-A and LoRA-B—are added. LoRA-A has dimensions E x d and d x L for Q, K, V, while LoRA-B has dimensions L x d and d x E for O projection, where d represents the LoRA dimension (rank). A scale factor s is introduced to control the impact of LoRA weights. In Eqs. 1 to 4, W denotes the projection weight matrix, and A and B represent the LoRA weights.

\begin{equation}
    Y^Q=QW^Q+Q(sA^QB^Q) 
\end{equation}
\begin{equation}
    Y^K=KW^K+K(sA^KB^K)
\end{equation}
\begin{equation}
    Y^V=VW^V+V(sA^VB^V)
\end{equation}
\begin{equation}
    Y^O=OW^O+O(sA^OB^O)
\end{equation}

\subsection{Multi-LoRA enablement on device}
An overview of the proposed framework for LoRA enablement and runtime switching is presented in Figure~\ref{fig_lora}. The framework has three approaches with various pros and cons. 

In the first one (see Figure ~\ref{fig_lora}a), different graphs are prepared for different LoRA enabled tasks and the weights of the layers where LoRA is not applied, are shared between the graphs. During runtime, the corresponding graph is triggered. This ensures saving a significant amount of memory.

The second approach (see Figure ~\ref{fig_lora}b) maintains a single graph with multiple LoRAs and facilitates dynamic switching between different tasks by applying one-hot encoded mask on the LoRAs. This approach provides inference benefit but increases the memory footprint.

The third approach (see Figure ~\ref{fig_lora}c) provides both latency and memory benefits by creating the foundation LLM graph with placeholders for LoRA weights and passing the LoRA weights as inputs along with tokens, assuming that LoRA weights for all use cases are of same dimensions. This simple yet elegant solution successfully enabled multiple LoRAs on device with low latency and memory.

\subsection{Model Optimizations for On-Device LLM}
Deploying Large Language Models (LLMs) on edge devices necessitates rigorous architectural and computational optimizations to meet the constraints of memory, latency, and power consumption while preserving functional accuracy. This section discusses several key model-level transformations and compilation-aware optimizations that enable efficient inference of LLMs on-device, tailored for modern NPUs and embedded GPUs.
\paragraph{Attention Parallelism.} Multi-head attention (MHA), a core component of transformer-based LLMs, presents challenges in hardware execution due to its sequential dependencies and shared projection layers. To facilitate parallel processing on hardware accelerators, we decompose the MHA into multiple single-head attention (SHA) operations. Each SHA can be independently mapped to a parallel core, improving throughput. 

This fine-grained decomposition also enables head-wise scheduling and execution on heterogeneous compute units, aligning with the architectural strengths of NPUs designed for independent low-dimensional matrix operations.

\paragraph{Linear-to-Convolutional Layer Transformation.} Another major optimization involves reinterpreting linear (dense) layers as 1x1 convolutions. While mathematically equivalent in functionality, this transformation allows the model to exploit highly optimized convolution kernels present in embedded hardware backends. Modern mobile NPUs and GPUs, originally optimized for vision tasks, offer superior support for convolutional operations via Winograd algorithms, tiling strategies, and pipelined execution units. By casting dense layers in the form of point-wise convolutions, we enable efficient reuse of existing high-performance operators, reducing both latency and energy overheads.

\paragraph{Constant Folding and Graph Simplification.}
During graph-level compilation, we apply constant folding to precompute operations involving static weights or configuration tensors. Examples include pre-multiplication of learned scale parameters in layer normalization or static reshaping and permutation of input dimensions. These algebraic simplifications reduce runtime computation and memory access, streamlining the execution graph. In combination with operator fusion—where adjacent linear and activation layers are merged into a single kernel—these techniques reduce kernel launch overhead and improve data locality in memory-constrained environments.

\paragraph{LLM Quantization}
Precision quantization plays a central role in enabling LLM inference on resource-constrained devices. In our framework, we employ Quantization-Aware Training (QAT) to simulate low-precision effects during training, allowing the model to learn robust representations under quantized conditions. We adopt a mixed-precision regime, where activations are quantized to per-tensor scaling for runtime simplicity to Int8 and weights per-channel to Int4, ensuring a trade-off between accuracy and compression. 

\begin{figure}[]
\centering
\includegraphics[width=0.5\textwidth]{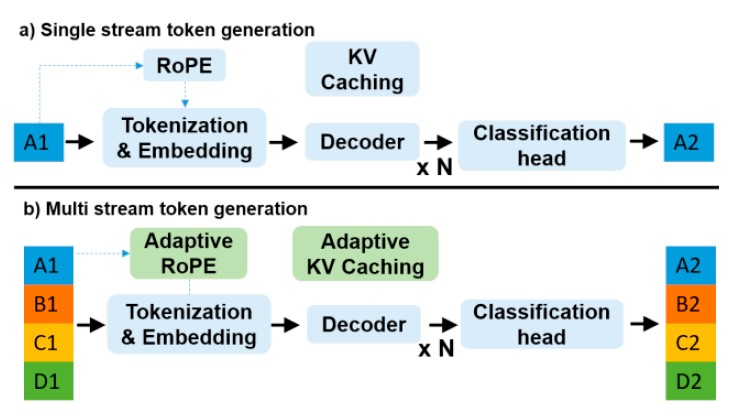}  
\captionsetup{justification=centering}
\caption{Schematic for Concurrent token generation}
\label{fig_smart_reply}
\end{figure}

\subsection{Concurrent token generation (CTG)}
To improve user experience, the LLM model generates multiple response options for personalization. For example, Smart-Reply provides 8 diverse prompts for users to choose from. However, this requires running the LoRA-enabled LLM sequentially 8 times, increasing latency by 8x. To address this, the proposed LoRA approaches enable handling multiple output streams simultaneously. The multi-stream token generation algorithm allows concurrent creation of n output streams, as shown in Figure ~\ref{fig_smart_reply}. More elaborate details on the flowsheet of on-device execution and the mask used for implementing the method are presented in the Appendix ~\ref{app:ctg}.

\subsection{Dynamic Self Speculative Decoding}
Speculative decoding has emerged as a promising approach to overcome inference-time inefficiencies in LLMs, which typically suffer from sequential, memory-bound decoding constraints. Some techniques use draft models and additional heads but they incur additional memory, while some early exit based methods require retraining the whole model to make the intermediate layers predict proper tokens. To avoid these issues, we utilized Prefix tuning based parameter efficient mechanism to fine-tune the foundational LLM, such that it generates multiple tokens in addition to the auto regressive token in one step.

In this method inspired from BiTA \citep{lin2025bita} task-specific m forecast embeddings are appended to the prompt embedding. These additional m forecast embedding ensure that the model predicts 1+m tokens at a time, thus converting its nature from auto-regressive to semi-auto-regressive (SAR). While the first token is sampled from the frozen LLM’s distribution, ensuring it to be same as the case without the forecast embedding, the remaining m tokens are generated with reduced fidelity, thus requiring verification during the inference. The fine-tuning procedure to generate SAR tokens is facilitated through the prefix tuning method (see Appendix ~\ref{app:ds2d} for further details).

\begin{figure*}[t!]
\centering
\includegraphics[width=\textwidth]{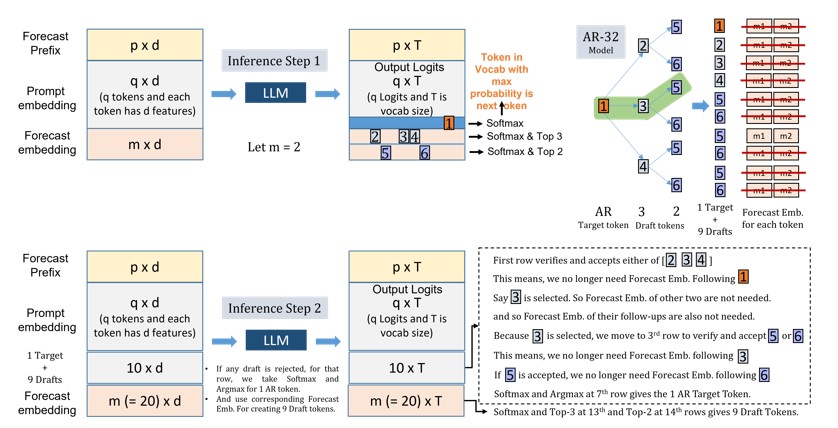}  
\captionsetup{justification=centering}
\caption{Approach for self-speculative decoding and the procedure for fine-tuning the model with Prefix tuning.}
\label{fig_ssd_infer2}
\end{figure*}

As illustrated in Figure  ~\ref{fig_ssd_infer2}, during the inference step 1, the forecast prefix (shape: p×d), prompt embeddings (shape: q×d), and forecast embeddings (shape: m × d) are passed as inputs to the LLM. In the output logits, the token marked in orange is generated from the prompt and is referred to as the "last verified token,". For the logits derived from the forecast embeddings, a softmax operation is performed, followed by sampling to generate multiple draft tokens. In the example provided in Figure  ~\ref{fig_ssd_infer2}, with m = 2, i.e., 2 forecast embeddings, (3, 2) branch configuration is implemented. 
In inference step 1, the top 3 and top 2 tokens are sampled from the logits of the first and second forecast embedding, respectively. These draft tokens are used to build a tree-based structure, as depicted in Figure ~\ref{fig_ssd_infer2}. In inference step 2, the generated draft tokens undergo verification along with new draft token generation. Verification is done by appending the 9 draft tokens to the prompt token, and new draft generation is done by appending a set of m forecast embeddings for each target and draft token. In this case, a total of 10 tokens and 20 forecast embeddings are passed as input. Based on the acceptance of the current draft tokens, the next draft tokens will be generated. In Figure ~\ref{fig_ssd_infer2}, since tokens 1-3-5 are accepted, drafts are generated from the output logits of forecast embeddings corresponding to token 5. The on-device deployment of this dynamic selection is facilitated through a mask described in the Appendix ~\ref{app:ds2d}.

\paragraph{Optimal Branch Configuration.} As shown in Figure ~\ref{fig_ssd_infer2}, the input shape (mainly rows) during inference (step 2) for a branch configuration of 3,2 corresponds to p+q+30 tokens. Since KV Cache mechanism will be implemented the p forecast prefix and all the prompt tokens except the previous token will be cached. Thus, the number of rows in input is 30. Since these inputs will be processed in parallel by the Transformer architecture, they are analogous to the batch size used during training or inference. Owing to the optimized architecture design of NPU (or GPUs), we consider the input size to be 32 (powers of 2) to ensure maximum benefit. Given the input size as 32, we can try different branch configurations that can result in a total of 31 tokens in the input + 1 previous token. Among all those configurations, the optimal configuration for use-case is determined as the one which results in maximum acceptance rate. Since, the branch configuration can be determined optimally for

%% file: section/4_experiments_methods.tex
\section{Experiments \& Methods}

\subsection{Experiments on GS24 with 1B LLM}

\begin{table}[h!]
\captionsetup{justification=centering, font=small}
\caption{On-Device Performance of LoRA weight sharing}
\label{tab:lora_weight_Sharing_1}
\centering
\small
\setlength{\tabcolsep}{1pt} 
\begin{tabular}{|l|c|c|c|c|}
\hline
 & \multicolumn{2}{c|}{\textbf{Base}} & \multicolumn{2}{c|}{\textbf{LoRA (2-tasks)}} \\
\cline{2-5}
 & \textbf{1st Token} & \textbf{Gen. model} & \textbf{1st Token} & \textbf{Gen. model} \\
\hline
\textbf{Num. graphs} & \multicolumn{2}{c|}{2} & \multicolumn{2}{c|}{4} \\
\hline
\textbf{Size} & \multicolumn{2}{c|}{718 MB} & \multicolumn{2}{c|}{1.1 GB} \\
\hline
\textbf{Latency} & 45ms & 22ms & 45ms & 22ms \\
\hline
\end{tabular}
\end{table}

\begin{table}[h]
\centering
\captionsetup{justification=centering}
\caption{On-Device Performance of LoRA weight masking and LoRA as input approaches measured on 2 tasks}
\label{tab:lora_weight_masking}
\resizebox{0.5\textwidth}{!}{%
\begin{tabular}{|l|l|c|c|c|c|}
\hline
\multirow{2}{*}{\textbf{Approach}} & \multirow{2}{*}{\textbf{Metric}} & \multicolumn{2}{c|}{\textbf{Prefill Mode}} & \multicolumn{2}{c|}{\textbf{Generative Mode}} \\ \cline{3-6}
 &  & \textbf{Base} & \textbf{LoRA} & \textbf{Base} & \textbf{LoRA} \\ \hline
\multirow{2}{*}{Masking} & Size & 718 MB & 894 MB & 718 MB & 894 MB \\ \cline{2-6}
 & Latency & 45ms & 75ms & 22ms & 30ms \\ \hline
\multirow{2}{*}{LoRA as Input} & Size & 718 MB & 686 MB & 718 MB & 686 MB \\ \cline{2-6}
 & Latency & 45ms & 52ms & 22ms & 25ms \\ \hline
\end{tabular}
}
\end{table}

\begin{table}[]
\captionsetup{justification=centering}
\centering
\caption{Inference time analysis with CTG}
\label{tab:smart_reply_gs24}
\resizebox{0.5\textwidth}{!}{%
\begin{tabular}{|c|p{2cm}|p{2cm}|p{2cm}|c|}
\hline
\textbf{Streams} & \textbf{Prefill Latency (ms)} & \textbf{AR Latency (ms)} & \textbf{Total Time (ms)} & \textbf{Formula} \\ \hline
1 & 40 & 23 & 174 & $(23 \times 8) + 40$ \\ \hline
8 & 40 & 23 & 63 & $23 + 40$ \\ \hline
\end{tabular}
}
\end{table}

After quantizing the model (see Appendix ~\ref{app:1b} for detailed results) and performing model optimization for its on-device deployment, the on-device performance was profiled on the GS24, with the proposed multi-LoRA method tested across the three approaches for enabling dynamic runtime switching. Results are presented in Tables \ref{tab:lora_weight_Sharing_1}, \ref{tab:lora_weight_masking}, showing that while all three approaches performed similarly for fewer use-cases (e.g., Table \ref{tab:lora_weight_masking}), the weight-as-input approach proved scalable for multiple use-cases.

\begin{table}[!h]
\centering
\captionsetup{justification=centering}
\caption{1B LLM On-Device Accuracy Results (\%) on GS24 Ultra.\vspace{0.2cm}}
\label{tab:g_eval_scores_1b_gs24}
\resizebox{0.48\textwidth}{!}{%
\begin{tabular}{|c|c|c|c|c|c|c|}
\hline
\textbf{Task} & \textbf{Korean} & \textbf{English} & \textbf{German} & \textbf{Spanish} & \textbf{French} & \textbf{Italian} \\ \hline
Correction & 96.83 & 97.60 & 90.20 & 95.90 & 95.37 & 94.58 \\ \hline
Style & 97.41 & 97.72 & 96.71 & 98.62 & 96.48 & 98.58 \\ \hline
Smart Reply & 103.40 & 111.13 & 97.36 & 96.02 & 97.37 & 100.79 \\ \hline
\end{tabular}
}
\end{table}

The accuracy of various tasks and languages measured on-device using the quantized LLM and LoRAs with proposed post-training quantization (PTQ) approaches is detailed in Table \ref{tab:g_eval_scores_1b_gs24}. 

We also evaluated the on-device accuracy of various tasks, such as correction, style, and Smart Reply, using the quantized LLM and LoRA with our proposed Post-Training Quantization (PTQ) approach. The results are presented in Table \ref{tab:g_eval_scores_1b_gs24}, which shows that our approach maintains accuracy while reducing latency.  The one-for-all foundation model’s overall memory and generation performance on-device is presented in Table ~\ref{tab:model-performance_1b}. Exhaustive results and demonstrations are provided in Appendix ~\ref{app:1b}.

\begin{table}[]
\captionsetup{justification=centering}
\centering
\caption{On-device Performance of One-for-all LLM model}
\label{tab:model-performance_1b}
\resizebox{0.4\textwidth}{!}{%
\begin{tabular}{|l|r|}
\hline
\multicolumn{2}{|c|}{\textbf{Memory and Initialization}} \\ \hline
Peak Memory (MB) & 967 \\
Load Time (ms) & 624.1 \\
First Token Latency (ms) & 155.9 \\
\hline
\multicolumn{2}{|c|}{\textbf{Generation Performance}} \\ \hline
Per-Token Latency (ms) & 24.19 \\
Token Generation Rate (tokens/sec) & 41.32 \\ \hline
Total End-to-End Time (sec) & 2.85 \\
\hline
\end{tabular}
}
\end{table}

\begin{table*}[!h]
\captionsetup{justification=centering}
\caption{3B LLM Performance for different use cases w/ \& w/o DS2D on GS25 Ultra}
\label{tab:use_cases_gs25}
\centering
\resizebox{0.9\textwidth}{!}{%
\begin{tabular}{|c|c|c|c|c|c|c|c|c|}
\hline
\multirow{2}{*}{\textbf{Use Case}} & \multirow{2}{*}{\textbf{Prefill (ms)}} & \multicolumn{2}{c|}{\textbf{Decode time (ms)}} & \multicolumn{2}{c|}{\textbf{Tokens/sec}} & \multirow{2}{*}{\textbf{Peak Mem (MB)}} & \multirow{2}{*}{\textbf{Init (ms)}} \\
\cline{3-6}
& &  \textbf{w/o DS2D} & \textbf{w/ DS2D} & \textbf{w/o DS2D} & \textbf{w/ DS2D} & &  \\
\hline
Correction & 208.72 & 50.17 & \textbf{22.3} & 19.93 & \textbf{44.84} & 2463.44 & 1507.96 \\
Composer & 209.57 & 53.23 & \textbf{28.57} & 18.79 & \textbf{35.00} & 2468.51 & 1499 \\
Style & 209.34 & 50.42 & \textbf{25.21} & 19.83 & \textbf{39.67} & 2465.42 & 1534.27 \\
Health & 210.7 & 51.92 & \textbf{28.77} & 19.26 & \textbf{34.76} & 2468.52 & 1551.98 \\
Summarization & 797.22 & 52.91 & \textbf{24.12} & 18.90 & \textbf{41.46} & 2486.98 & 1576.78 \\
Natural Language & 209.8 & 51.87 & \textbf{23.64} & 19.28 & \textbf{42.30} & 2480.09 & 1547.54 \\
ST Energy & 210.77 & 52.76 & \textbf{33.48} & 18.95 & \textbf{29.87} & 2486.3 & 1487.4 \\
AI Brief & 404.88 & 52.85 & \textbf{37.75} & 18.92 & \textbf{26.49} & 2484.68 & 1509.68 \\
\hline
\end{tabular}
}
\end{table*}


\subsection{Scaling to a 3B LLM on GS25 with DS2D}

With the deployment of a 3B LLM on GS25, we further accelerated the model using our proposed Dynamic Self-Speculative Decoding (DS2D) approach. The implementation of DS2D, which chooses the optimal branch for each use-case, resulted in a 2-fold improvement in tokens per second compared to non-speculative decoding, as shown in Table \ref{tab:use_cases_gs25}. Table ~\ref{tab:ds2d_toks/sec_gs25} presents the acceleration obtained with DS2D measured in average tokens per inference \& tokens per second on GS25 Ultra, with optimal branch configurations enabling a single graph for all use-cases. On-device accuracy comparisons between quantized and full-precision models across 9 languages
are presented in Table ~\ref{tab:G_Eval_GS25}. This improvement led to the real-world deployment of our approach in the Galaxy S25 family. Additional results included in appendix ~\ref{app:3b}.

\paragraph{Appendix}Due to space constraints, the following details, extended experiments, and on-device demonstrations are provided in the appendix:
    \begin{itemize}
        \item \textbf{Concurrent token generation (CTG):} Pipeline and mask for Concurrent token generation.
        \item \textbf{Dynamic Self Speculative Decoding:} Details of finetuning LLM for SAR generation and mask used during inference.
        \item \textbf{Deploying a 1B LLM:} Comparison between full precision and quantized models, latency improvements with proposed graph optimizations
        and on device demonstrations of smart reply and style suggestion usecases.
        \item \textbf{Scaling to a 3B LLM:} Demonstrations on device for writing assist, summarization and health energy score use cases.
    \end{itemize}




\begin{table*}[h!]
\captionsetup{justification=centering}
\caption{Acceleration obtained in 8 different use-cases under different branch configurations on GS25 with SM8750 Qualcomm chipset}
\label{tab:ds2d_toks/sec_gs25}
\centering
\resizebox{\textwidth}{!}{%
\begin{tabular}{|c|c|c|c|c|c|c|c|c|c|c|c|c|c|c|c|c|}
\toprule \hline
\multirow{2}{*}{\textbf{Branch Config}} &
\multicolumn{2}{c|}{\textbf{Correction}} & 
\multicolumn{2}{c|}{\textbf{Composer}} & 
\multicolumn{2}{c|}{\textbf{Style}} & 
\multicolumn{2}{c|}{\textbf{Health}} & 
\multicolumn{2}{c|}{\textbf{Summarization}} & 
\multicolumn{2}{c|}{\textbf{Gallery}} & 
\multicolumn{2}{c|}{\textbf{ST Energy}} & 
\multicolumn{2}{c|}{\textbf{AI Brief}}\\ \cline{2-17}
& tokens/inf & tokens/s & tokens/inf & tokens/s & tokens/inf & tokens/s & tokens/inf & tokens/s & tokens/inf & tokens/s & tokens/inf & tokens/s & tokens/inf & tokens/s & tokens/inf & tokens/s \\ \hline
\midrule
15 & 1.7 & 39.26 & \textbf{1.72} & \textbf{40.35} & 1.7 & 39.27 & \textbf{1.54} & \textbf{36.05} & 1.76 & 40.82 & \textbf{1.59} & \textbf{36.84} & \textbf{1.45} & \textbf{33.88} & \textbf{1.57} & \textbf{36.46} \\ \hline
1,8 & 2.05 & 45.88 & 1.51 & 35.47 & 1.81 & 41.38 & 1.31 & 30.69 & 1.7 & 39.07 & 1.49 & 34.45 & 1.08 & 25.49 & 1.32 & 30.91 \\ \hline
2,3 & 2.14 & 47.69 & 1.56 & 36.57 & 1.94 & 43.85 & 1.37 & 31.98 & 1.74 & 39.8 & 1.63 & 37.32 & 1.21 & 28.51 & 1.45 & 33.44 \\ \hline
3,2 & 2.13 & 47.94 & 1.6 & 37.61 & \textbf{1.96} & \textbf{44.47} & 1.41 & 33.09 & \textbf{1.8} & \textbf{41.34} & 1.58 & 36.38 & 1.25 & 29.35 & 1.5 & 34.74 \\ \hline
4,1 & 2.02 & 45.31 & 1.62 & 37.83 & 1.85 & 42.05 & 1.48 & 34.5 & 1.77 & 40.37 & 1.52 & 35.15 & 1.3 & 30.31 & 1.48 & 34.09 \\ \hline
1,1,5 & 2.25 & 49.44 & 1.39 & 32.32 & 1.84 & 41.51 & 1.24 & 28.99 & 1.58 & 35.98 & 1.35 & 31.27 & 1.07 & 25.13 & 1.27 & 29.33 \\ \hline
1,2,2 & \textbf{2.27} & \textbf{49.6} & 1.43 & 33.24 & 1.92 & 43.13 & 1.25 & 29.1 & 1.67 & 37.98 & 1.43 & 32.86 & 1.07 & 25.02 & 1.31 & 30.25 \\ \hline
2,1,1 & 2 & 44.55 & 1.49 & 34.36 & 1.8 & 40.58 & 1.32 & 30.48 & 1.64 & 37.43 & 1.46 & 33.39 & 1.21 & 27.92 & 1.43 & 32.92 \\ \hline
1,1,1,2 & 2.14 & 46.95 & 1.37 & 31.57 & 1.81 & 40.47 & 1.24 & 28.69 & 1.53 & 33.82 & 1.35 & 29.56 & 1.07 & 23.87 & 1.26 & 28.39 \\ \hline
\bottomrule
\end{tabular}
}
\end{table*}

\begin{table*}[h!]
\centering
\caption{On-Device Accuracy measured for different use-cases in 9 languages on GS25. G-Eval measured with ChatGPT 4.5 as Evaluator Model and Relative Accuracy (RA) = $100 \times \frac{\text{OnDevice QAT Accuracy}}{\text{FP32 Accuracy}}$}
\label{tab:G_Eval_GS25}
\begin{tabular}{|l|l|c|c|c|c|c|c|}
\hline
\multirow{2}{*}{\textbf{Precision}} & \multirow{2}{*}{\textbf{Task/Language}} & \multicolumn{2}{c|}{\textbf{Korean}} & \multicolumn{2}{c|}{\textbf{English}} & \multicolumn{2}{c|}{\textbf{German}} \\ \cline{3-8}
 &  & \textbf{G-Eval} & \textbf{RA} & \textbf{G-Eval} & \textbf{RA} & \textbf{G-Eval} & \textbf{RA} \\ \hline
Fp32 & \multirow{2}{*}{Correction} & 2.53 & NA & 2.664 & NA & 2.35 & NA \\ \cline{1-1} \cline{3-8}
QAT - On Device &  & 2.52 & 99.6047 & 2.7424 & 102.943 & 2.3892 & 101.668 \\ \hline
Fp32 & \multirow{2}{*}{Style} & 2.94 & NA & 2.935 & NA & 2.806 & NA \\ \cline{1-1} \cline{3-8}
QAT - On Device &  & 2.96 & 100.51 & 2.928 & 99.7445 & 2.871 & 102.309 \\ \hline
Fp32 & \multirow{2}{*}{Smart Reply} & 2.7 & NA & 2.834 & NA & 2.621 & NA \\ \cline{1-1} \cline{3-8}
QAT - On Device & & 2.67 & 99.0724 & 2.81 & 99.1531 & 2.56 & 97.6726 \\ \hline
Fp32 & \multirow{2}{*}{Summarization} & 2.71 & NA & 2.65 & NA & 2.65 & NA \\ \cline{1-1} \cline{3-8}
QAT - On Device &  & 2.693 & 99.3542 & 2.688 & 101.446 & 2.641 & 99.66031 \\ \hline
\hline
\multirow{2}{*}{} & \multirow{2}{*}{} & \multicolumn{2}{c|}{\textbf{Spanish}} & \multicolumn{2}{c|}{\textbf{French}} & \multicolumn{2}{c|}{\textbf{Italian}} \\ \cline{3-8}
 &  & \textbf{G-Eval} & \textbf{RA} & \textbf{G-Eval} & \textbf{RA} & \textbf{G-Eval} & \textbf{RA} \\ \hline
Fp32 & \multirow{2}{*}{Correction} & 2.48 & NA & 2.64 & NA & 2.715 & NA \\ \cline{1-1} \cline{3-8}
QAT - On Device &  & 2.535 & 102.22 & 2.608 & 98.769 & 2.635 & 97.053 \\ \hline
Fp32 & \multirow{2}{*}{Style} & 2.958 & NA & 2.838 & NA & 2.743 & NA \\ \cline{1-1} \cline{3-8}
QAT - On Device &  & 2.891 & 97.755 & 2.869 & 101.103 & 2.807 & 102.34 \\ \hline
Fp32 & \multirow{2}{*}{Smart Reply} & 2.675 & NA & 2.595 & NA & 2.588 & NA \\ \cline{1-1} \cline{3-8}
QAT - On Device & & 2.65 & 99.065 & 2.68 & 103.276 & 2.63 & 101.62 \\ \hline
Fp32 & \multirow{2}{*}{Summarization} & 2.647 & NA & 2.678 & NA & 2.502 & NA \\ \cline{1-1} \cline{3-8}
QAT - On Device &  & 2.653 & 100.227 & 2.731 & 101.96 & 2.421 & 96.7619 \\ \hline
\hline
\multirow{2}{*}{} & \multirow{2}{*}{} & \multicolumn{2}{c|}{\textbf{Portuguese}} & \multicolumn{2}{c|}{\textbf{Chinese}} & \multicolumn{2}{c|}{\textbf{Japanese}} \\ \cline{3-8}
 &  & \textbf{G-Eval} & \textbf{RA} & \textbf{G-Eval} & \textbf{RA} & \textbf{G-Eval} & \textbf{RA} \\ \hline
Fp32 & \multirow{2}{*}{Correction} & 2.418 & NA & 2.24 & NA & 2.385 & NA \\ \cline{1-1} \cline{3-8}
QAT - On Device &  & 2.638 & 109.078 & 2.174 & 97.031 & 2.305 & 96.6457 \\ \hline
Fp32 & \multirow{2}{*}{Style} & 2.783 & NA & 2.929 & NA & 2.928 & NA \\ \cline{1-1} \cline{3-8}
QAT - On Device &  & 2.836 & 101.919 & 2.878 & 98.228 & 2.91 & 99.392 \\ \hline
Fp32 & \multirow{2}{*}{Smart Reply} & 2.534 & NA & 2.735 & NA & 2.534 & NA \\ \cline{1-1} \cline{3-8}
QAT - On Device & & 2.67 & 105.367 & 2.76 & 100.91 & 2.25 & 88.7924 \\ \hline
Fp32 & \multirow{2}{*}{Summarization} & 2.647 & NA & 2.64 & NA & 2.646 & NA \\ \cline{1-1} \cline{3-8}
QAT - On Device &  & 2.624 & 99.13109 & 2.189 & 82.9167 & 2.535 & 95.8042 \\ \hline
\end{tabular}
\end{table*}


%% file: section/5_conclusion.tex
\section{Conclusion}

This work presents an effort to optimize the inference pipeline for deploying a single pre-trained foundation model to serve multiple Generative AI use-cases in the Galaxy AI feature of Samsung flagships. The proposed approach implements methods and optimizations for enabling multiple quantized low rank adapters capable of switching on demand at runtime. The empirical analysis demonstrates 8 different uses-cases with a single large language foundation model that results in 3 and 2 fold improvement in on-device memory and token rate, respectively.  

The framework proposed in this research is first-of-its-kind which led to the successful on-device deployment of Generative AI-based language understanding use-cases. This framework will serve as a guide for best practices in optimizing the memory-intensive foundation models for efficient on-device deployment. The concepts such as the quantization, beyond the foundation models for Generative AI and apply to all kinds of AI models, resulting in a much larger scope, and efficient and sustainable on-device AI.  

This work establishes a scalable and efficient approach for deploying LLMs fully on-device. Through model re-engineering, multi-LoRA handling and the development of novel Self-Speculative Decoding, we achieve real-time inference across multiple tasks and languages on mobile NPUs. Our framework successfully commercialized advanced LLM capabilities without cloud reliance, setting a new benchmark for edge AI.

%% file: section/6_limitations.tex
\section{Limitations}
While our system achieves significant improvements in memory efficiency, latency, and multi-use-case deployment of LLMs on-device, it is subject to limitations which will be addressed in future works:
\begin{itemize}
    \item 
    LoRA Dimensions: Our LoRA-as-input design assumes that all LoRAs have identical dimensionality. This is necessary to maintain compatibility with the frozen inference graph and its fixed placeholder structure. Supporting heterogeneous LoRA sizes would require dynamic graph construction or runtime shape management, which is infeasible under current deployment 
    constraints.
    \item
    Forecast Embeddings Are Static in DS2D: In our Dynamic Self-Speculative Decoding (DS2D) approach, the forecast embeddings used to precompute semi-autoregressive (SAR) tokens remain static across inference steps. Since they are not conditioned on the evolving input sequence, this can lower the token acceptance rate, particularly for inputs with high semantic variance or long-term dependencies.
    \item 
    Predefined Style Variants: The multi-stream decoding strategy relies on a fixed set of stylistic variants defined at compile time. While this enables concurrent generation, it does not support dynamically defined or user-specified styles without recompilation or memory duplication.
\end{itemize}

%% file: section/7_appendix.tex
\appendix

\section{Appendix}
\label{sec:appendix}
\subsection{Concurrent token generation (CTG)}\label{app:ctg}
\begin{figure*}[!h]
\centering
\includegraphics[width=\textwidth]{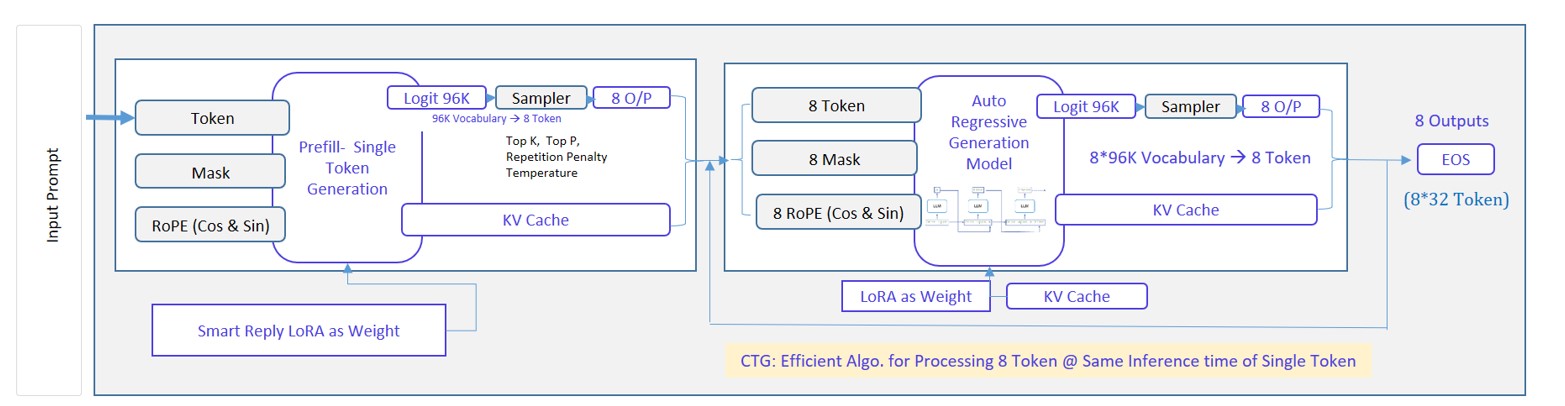}  
\captionsetup{justification=centering}
\caption{Flowsheet for the execution of Concurrent token generation (CTG) on device}
\label{fig_smart_reply_pipeline}
\end{figure*}

\begin{figure*}[!h]
\centering
\includegraphics[width=\textwidth]{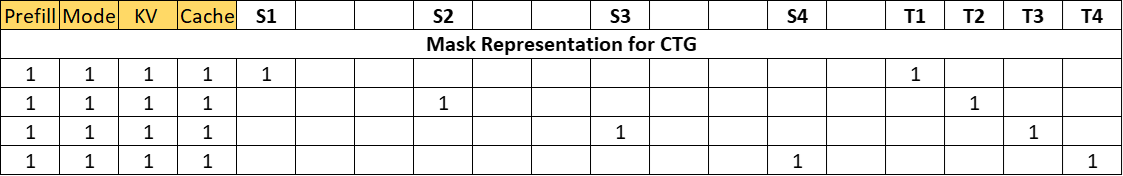}  
\captionsetup{justification=centering}
\caption{Mask used during CTG. KV Cache is divided into 5 parts: one for the common prefill cache and four for individual sentences (S1 - S4). The tokens (T1 - T4) attend to the prefill and their respective sentence tokens.All empty values are 0's}
\label{fig_smart_reply_mask}
\end{figure*}

Figure ~\ref{fig_smart_reply_pipeline} illustrates the pipeline for concurrent token generation. It starts with a single Prefill model execution, followed by a sampler generating 8 distinct output tokens. These tokens trigger the generative model flow for 8 sentences. An 8-fold generative model processes 8 inputs simultaneously, yielding 8 outputs. The KV-cache is divided into 8 isolated segments, with masks ensuring each token stream interacts only with its Prefill context and corresponding segment (see Figure ~\ref{fig_smart_reply_mask}).

\subsection{Dynamic Self Speculative Decoding}\label{app:ds2d}
\begin{figure*}[!h]
\centering
\includegraphics[width=\textwidth]{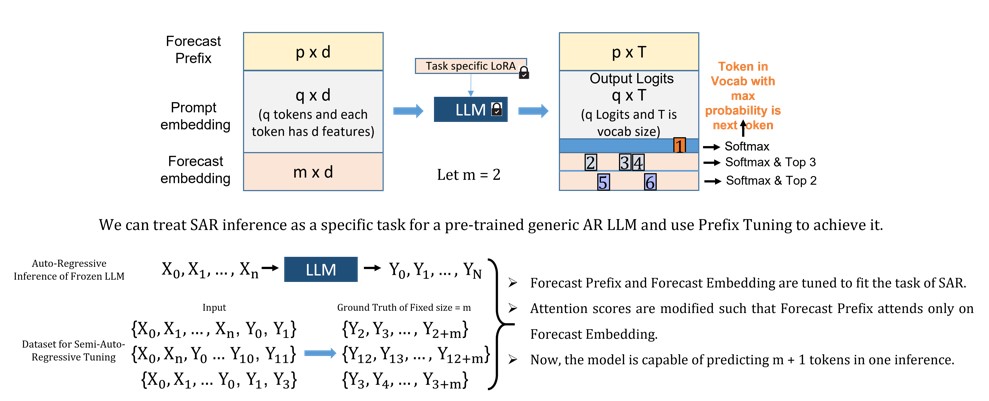}  
\captionsetup{justification=centering}
\caption{Approach for self-speculative decoding and the procedure for fine-tuning the model with Prefix tuning.}
\label{fig_ssd_training}
\end{figure*}

\begin{figure*}[!h]
\centering
\includegraphics[width=\textwidth]{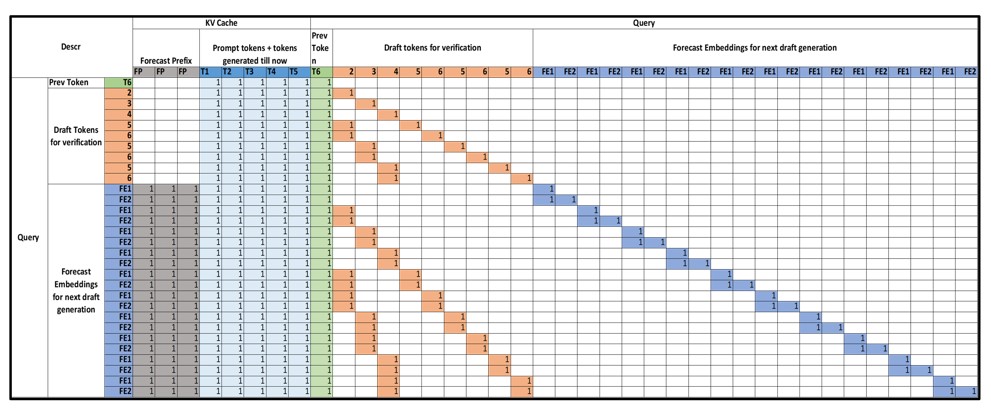}  
\captionsetup{justification=centering}
\caption{The mask used during inference step 2 in Figure 6. All the empty values indicate 0s. }
\label{fig_ssd_mask}
\end{figure*}

 To enable the LLM to generate tokens semi-auto-regressively, Prefix Tuning is employed. The dataset can be generated from the frozen pre-trained LLM in auto-regressive manner and can be rearranged to facilitate the finetuning as shown in Figure  ~\ref{fig_ssd_training}. The forecast prefix induces the capacity of semi-auto regression into already trained and frozen LLM. Further the causal mask is modified to avoid the interaction of the forecast prefix with prompt token as shown in Figure ~\ref{fig_ssd_mask}. Since the m new tokens are of low-fidelity, these tokens need to be validated during the next inference step, while generating new draft tokens simultaneously.

\paragraph{Optimal Branch Configuration}The configuration (3, 2) which results in 9 draft tokens, is dynamic and use-case specific, allowing the determination of optimal branching for each use-case. The dynamic selection is learnt during the fine-tuning step and facilitated during the inference step through the mask as shown in Figure ~\ref{fig_ssd_mask}. Note that, since the first inference step does not have to validate any draft tokens, the mask used in inference step 1 will be much simpler than the one described in ~\ref{fig_ssd_mask}.

\subsection{Experiments}

\subsubsection{Deploying a 1B LLM on GS24}\label{app:1b}The first set of experiments focuses on deploying a 1B-parameter LLM on the GS24, utilizing the sm8650 chipset’s Neural Processing Unit (NPU). The study spans 36 use-cases across 4 tasks—text style/tone transfer, spelling/grammar correction, smart reply, and health summarization—in 9 languages: Korean, English, German, Spanish, Italian, French, Dutch, Chinese, and Japanese. Four distinct LoRAs were developed for these use-cases, and the results were evaluated after quantization, optimization, and deployment on-device.


\begin{table}[h]
\captionsetup{justification=centering}
\caption{Memory comparison between full precision and quantized models with LoRA for 4 tasks}
\label{tab:memory_Comparison}
\centering
\resizebox{0.4\textwidth}{!}{%
\begin{tabular}{|c|c|}
\hline
\textbf{Model} & \textbf{ROM (MB)} \\
\hline
FP16 LLM + LoRA & 1800+120 \\
\hline
INT4 LLM + LoRA & 600+120 \\
\hline
\end{tabular}
}
\end{table}

 Memory comparisons between full-precision models and their quantized variants (with weights and activations in INT-4/INT-16 and LoRA weights in INT-16) are shown in Table \ref{tab:memory_Comparison} highlighting a 3x compression benefit from quantization. Graph optimizations, including scalar folding and LoRA multi-head splitting, are detailed in Table \ref{tab:performance_improvement_1}, with the latter improving quantization accuracy without affecting latency. The one-for-all foundation model’s on-device performance is presented in Table \ref{tab:model-performance_1b}.

\begin{table}[]
\captionsetup{justification=centering}
\caption{Performance improvement obtained with proposed Graph optimizations}
\label{tab:performance_improvement_1}
\centering
\resizebox{0.5\textwidth}{!}{%
\begin{tabular}{|l|c|c|}
\hline
\multirow{2}{*}{\textbf{Model-Type}} & \multicolumn{2}{c|}{\textbf{Latency (ms)}} \\ \cline{2-3}
 & \textbf{First token} & \textbf{Generative model} \\ \hline
Scalar-folding & 45.8 & 19.385 \\ \hline
Without Scalar-folding & 47.3 & 20.51 \\ \hline
LoRA-B splitting + Scalar folding & 45.9 & 19.45 \\ \hline
LoRA-B composite + Scalar folding & 45.8 & 19.385 \\ \hline
K-untransposed & 52 & 23 \\ \hline
K-transposed & 45.8 & 19.385 \\ \hline
\end{tabular}
}
\end{table}


\begin{figure*}[h!]
\centering
\includegraphics[width=\textwidth]{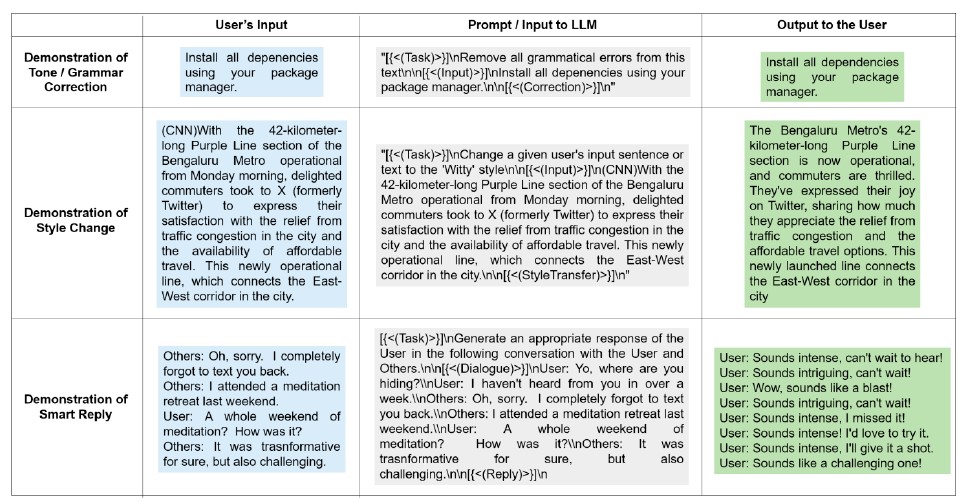}  
\captionsetup{justification=centering}
\caption{Demonstration of User’s inputs, corresponding prompts to the LLM, and the outputs generated by the LLM model with deployment of corresponding LoRAs.}
\label{fig_gs24_llm_demo}
\end{figure*}


\begin{figure*}[h!]
\centering
\includegraphics[width=\textwidth]{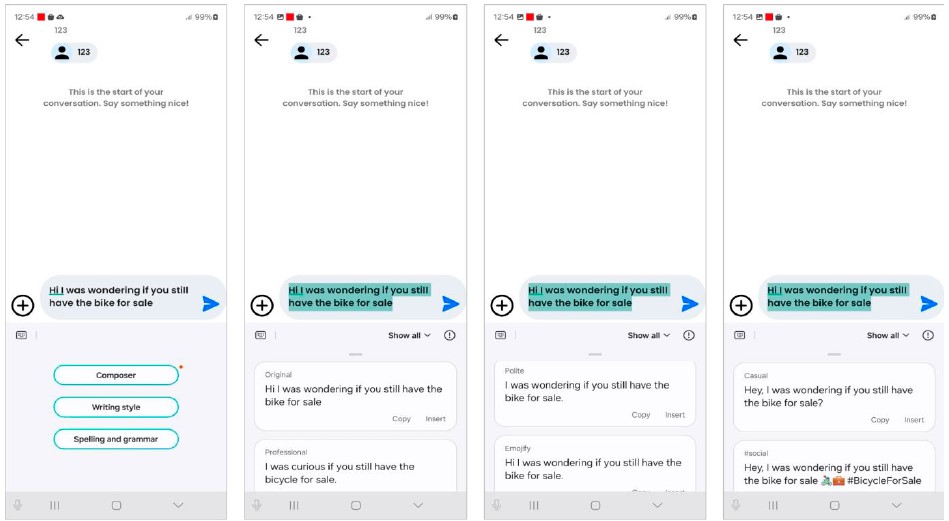}  
\captionsetup{justification=centering}
\caption{Demonstration of Style suggestion (Professional, Polite, Emojify, Casual \& Social) on GS24 Ultra.}
\label{fig_gs24_style_Demo}
\end{figure*}

\begin{figure*}[h!]
\centering
\includegraphics[width=\textwidth]{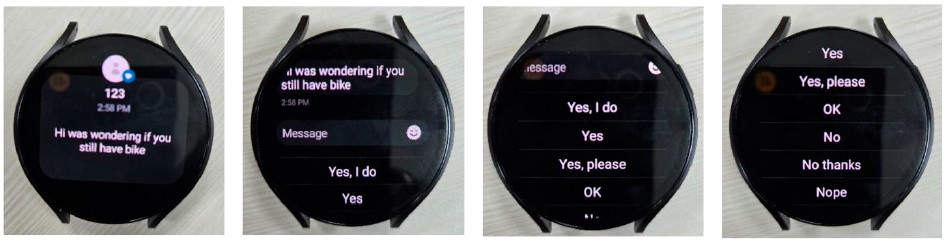}  
\captionsetup{justification=centering}
\caption{Demonstration of Smart Reply in Samsung Galaxy Smart Watch}
\label{fig_gs24_watch_Demo}
\end{figure*}

The demonstration of User’s inputs, corresponding prompts to the LLM, and the outputs generated by the LLM model with deployment of corresponding LoRAs is illustrated in Figure \ref{fig_gs24_llm_demo}. The on-device user inference for different tasks and languages was demonstrated on the GS24 and extended to the Samsung Galaxy Smartwatch via smart connectivity (Figures \ref{fig_gs24_style_Demo} \& \ref{fig_gs24_watch_Demo}), showcasing the method’s applicability across devices.

\subsubsection{Scaling to a 3B LLM with DS2D}\label{app:3b} Building on the success of the 1B LLM deployment, the second set of experiments scales up to a 3B-parameter model deployed on the GS25 Ultra, leveraging DS2D for further acceleration. The proposed method was applied to all 8 use-cases of language understanding commercialized in GS25 (Qualcomm) and planned for Galaxy Flip/Fold 7 (LSI-Exynos). 

\begin{figure*}[h!]
\centering
\includegraphics[width=\textwidth]{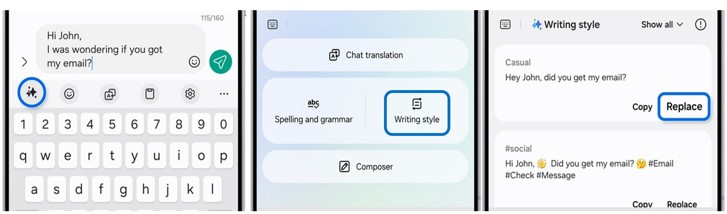}  
\captionsetup{justification=centering}
\caption{Demonstration of writing assist usecase in chat on Samsung GS25 Ultra Smart Phone }
\label{gs25_fig_10}
\end{figure*}

\begin{figure*}[h!]
\centering
\includegraphics[width=\textwidth]{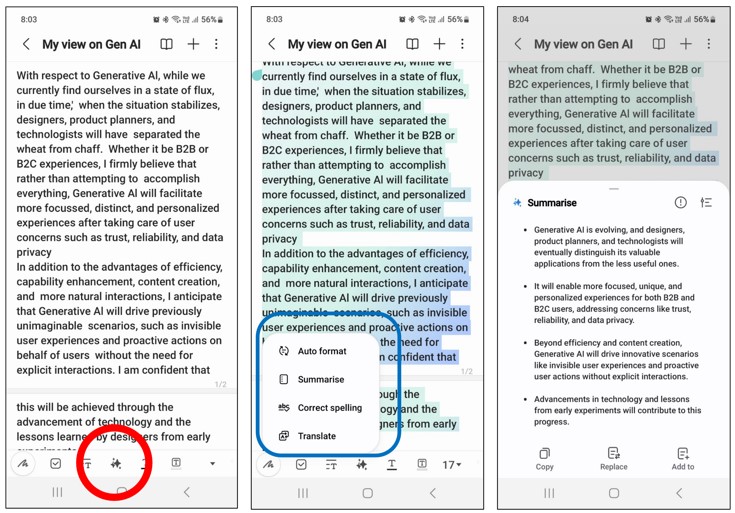}  
\captionsetup{justification=centering}
\caption{Demonstration of summarization usecase on Samsung GS25 Ultra Smart Phone }
\label{gs25_fig_8}
\end{figure*}

\begin{figure*}[h!]
\centering
\includegraphics[width=\textwidth]{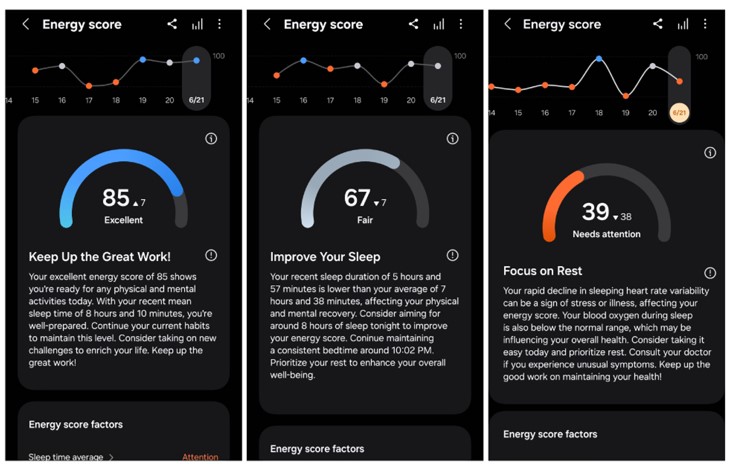}  
\captionsetup{justification=centering}
\caption{Demonstration of Energy score usecase in health application on Samsung GS25 Ultra Smart Phone }
\label{gs25_fig_9}
\end{figure*}

Post-deployment demonstrations of LLM applications for writing assist, summarization and health energy score, and  are shown in Figures \ref{gs25_fig_10} \ref{gs25_fig_8} \& \ref{gs25_fig_9}   highlighting the method’s real-world applicability.

